\documentstyle[aps]{revtex} 
\newtheorem{theorem}{Theorem}[section]
\newtheorem{corollary}[theorem]{Corollary} 
\newtheorem{proposition}[theorem]{Proposition}
\newtheorem{lemma}[theorem]{Lemma}  
\newtheorem{remark}[theorem]{Remark}
  
\newcommand{\proof}{\noindent {\bf Proof. }} 
\newcommand{\qed}{\hfill $\Box$ \vskip 2ex} 
\newfont{\BB}{msbm10} \def\R{\mbox{\BB R}} \def\C{\mbox{\BB C}}
\newcounter{acount}

\begin{document} 
\title{Stochastic mechanics and the Feynman integral}
\author{Michele Pavon}
\address{Dipartimento di  Elettronica e
Informatica, Universit\`a di  Padova, via Gradenigo 6/A,
and LADSEB, CNR,
35131 Padova, Italy,  Electronic mail: {\em pavon@dei.unipd.it}}
\date{\today}
\maketitle
\begin{abstract}
The Feynman integral is given a stochastic interpretation in
the framework of Nelson's stochastic mechanics employing a time-symmetric
variant of Nelson's kinematics recently developed by the author. 
\end{abstract}
\noindent
{\bf Running Title:} Stochastic mechanics and Feynman integral\\
{\bf PACS number:} 03.65.Bz
\section{Introduction} In 1964, Nelson, exploiting results of Kato and Trotter,
established the following important result \cite{N64}.
\begin{theorem}\label{th1} Let $V$ be a real function on
$\R^n$ belonging to the Kato class, let $\psi_0\in{\cal L}^2(\R^n)$, and let 
${\bf H} = -\frac{\hbar^2}{2m}\Delta + V(x)$ be the {\it Hamiltonian operator}.
Then, with
$x=x_0$,

\begin{eqnarray}\nonumber
\psi(t,x):&=&\left(\exp [-\frac{i}{\hbar}t{\bf H}]\psi_0\right)(x)={\cal
L}^2-\lim_{l\rightarrow\infty}\left[\frac{2\pi\hbar
it}{lm}\right]^{-\frac{nl}{2}}\\&&\int\cdots\int\exp\left[-\sum_{j=1}^{l}
\frac{i}{\hbar}\left(-\frac{m}{2}\frac{|x_j-x_{j-1}|^2}
{t/l}+V(x_j)\frac{t}{l}\right)\right]\psi_0(x_l)dx_l\cdots dx_1 .\nonumber
\end{eqnarray} 

\end{theorem} This result gives a precise meaning to the Feynman integral
\cite{F1}. There exists, by now, a large body of literature investigating
various aspects of the Feynman integral and its generalization, see
\cite{F2}-\cite{CDW} and references therein. Two years later, Nelson,
elaborating on previous work of F\'enyes and others, laid the foundations of a
quantization procedure for classical dynamical systems based on diffusion
processes \cite{N0}. The purpose of this paper is to show that there is a
connection between \cite{N64} and \cite{N0}.  More explicitly, we shall exhibit
a natural interpretation of Theorem \ref{th1} within Nelson's stochastic
mechanics
\cite{N0}-\cite{BCZ}. 

As is well known, a close formal analogy between Feynman and Wiener integrals
was observed very early. In order to emphasize the crucial difficulty in making
this analogy complete, we recall a few well known facts. Let us consider the
free case
$V\equiv 0$. Then, 
\begin{equation}\label{FREE}\psi(t,x):=\left(\exp [-\frac{i}{\hbar}t{\bf
H}]\psi_0\right)(x)=\left[\frac{2\pi\hbar
it}{m}\right]^{-\frac{n}{2}}\int\exp\left[-
\frac{i}{\hbar}\left(-\frac{m}{2\hbar}\frac{|x-y|^2}
{t}\right)\right]\psi_0(y)dy,
\end{equation} since

\begin{equation}\label{1} K(s,y,t,x):=\left[\frac{2\pi\hbar
i(t-s)}{m}\right]^{-\frac{n}{2}}\exp\left[\frac{im}{2\hbar}\frac{|x-y|^2}
{t-s}\right]
\end{equation} is the fundamental solution of

$$\frac{\partial{\psi}}{\partial{t}} = \frac{i\hbar}{2m}\Delta\psi.
$$ Consider the heat equation

\begin{equation}\label{2}
\frac{\partial{u}}{\partial{t}} =
\frac{1}{2}\Delta u,
\end{equation} whose fundamental solution is
\begin{equation}\label{3} p(s,y,t,x):=\left[2\pi(t-s)\right]
^{-\frac{n}{2}}\exp\left[-\frac{|x-y|^2} {2(t-s)}\right],\quad s<t.
\end{equation}  The solution of (\ref{2}), with initial condition $u_0$ at time
$t=0$, is then given by
$$u(t,x)=\int p(0,y,t,x)u_0(y)dy.
$$ On the other hand, $p(s,y,t,x)$ is also the transition density of a standard,
$n$-dimensional Wiener process $W$. Hence, we immediately get the probabilistic
representation

\begin{equation}\label{4} u(t,x)=E\left\{u_0(W(0))|W(t)=x\right\}.
\end{equation} Moreover, the kernel (\ref{3}) may be employed to construct
Wiener measure on path space via the Riesz-Markov representation theorem
\cite{N64}. Formula (\ref{4}) may be then replaced by
\begin{equation}\label{5} u(t,x)=\int_{\Omega}u_0(\omega(0))d{\cal
W}_{tx}(\omega),
\end{equation} where $\Omega:=C([0,t];\R^n)$. With the help of the Trotter
product formula, it is then possible to derive the Feynman-Kac formula for the
semigroup $\exp [-t(-\frac{1}{2}\Delta+V)]$
\cite{N64}.\\
\noindent In 1956, Gelfand and Yaglom suggested that the same route could
be followed in order to give sense to the Feynman integral as a path-integral
\cite{GY}. However, as argued by Cameron \cite{C}, kernel (\ref{1}) cannot be
employed to construct a countably additive path-space measure. In particular,
even in the free case $V\equiv 0$, and differently from the diffusion case,
there is no probabilistic interpretation of formula (\ref{FREE}), as we don't
have a probabilistic interpretation of kernel (\ref{1}). 

In this paper, we show that a probabilistic interpretation of (\ref{1}) is
possible in the framework of Nelson's stochastic mechanics. More explicitly, it
is possible to connect the kernel (\ref{1}) to the bi-directional generator
$L_b$ of the Nelson process (Proposition \ref{pro2}) very much the same way that
the kernel (\ref{3}) is connected to the usual generator of the Markov process
in the diffusion case (Proposition \ref{pro7}). The bi-directional generator of
the Nelson process ( see (\ref{GBD}) for the definition) originates from a
certain time-symmetric differential for finite-energy diffusions that has been
used in
\cite{P1}-\cite{P3} to develop elements of Lagrangian and Hamiltonian dynamics
within Nelson's stochastic mechanics. Moreover, as we showed in \cite{P4}, this
time-symmetric kinematics permits to {\em derive} the collapse of the
wave function after a position measurement through a stochastic variational
principle. The connection between the operators
$(\frac{\partial}{\partial{t}}+L_b)$ and
$(\frac{\partial}{\partial{t}}+\frac{i}{\hbar}{\bf H})$, where ${\bf H}$ is the
Hamiltonian operator, is given in Theorem \ref{th2}. The latter generalizes a
well-known unitary correspondence between the usual generator and the Hamiltonian
operator through the so-called {\em ground state transformation}. 

\section{Nelson-F\"{o}llmer kinematics of finite-energy diffusions}

In this section, we review some basic results of the kinematics of diffusion
processes. More information and the proofs may be found in \cite{N1}-\cite{N2},
\cite{F} and -\cite{KS}. Let
$(\Omega,{\cal E},{\bf P})$ be a probability space. A stochastic process
$\{\xi(t);t_0\le t\le t_1\}$ mapping
$[t_0,t_1]$ into
$L^2_n(\Omega,{\cal E},{\bf P})$ is called a {\it finite-energy diffusion} with
constant diffusion coefficient $I_n\sigma^2$ if the increments admit the
representation

\begin{equation}\label{K1} \xi(t)-\xi(s)=\int_s^t\beta(\tau)d\tau+\sigma
[w_+(t)-w_+(s)],\quad t_0\le s<t\le t_1, \end{equation} where the {\it forward
drift} $\beta(t)$ is at each time $t$ a measurable function of the past
$\{\xi(\tau);0\le \tau\le t\}$, and $w_+(\cdot)$ is a standard, n-dimensional
{\it Wiener process} with the property that $w_+(t)-w_+(s)$ is independent of
$\{\xi(\tau);0\le \tau\le s\}$. Moreover, $\beta$ must satisfy the finite-energy
condition
\begin{equation}\label{K2}
E\left\{\int_{t_0}^{t_1}\beta(t)\cdot\beta(t)dt\right\}<\infty.
\end{equation} In \cite{F}, F\"{o}llmer has shown that a finite-energy diffusion
also admits a reverse-time differential. Namely, there exists a measurable
function $\gamma(t)$ of the future
$\{\xi(\tau);t\le \tau\le t_1\}$, called {\it backward drift}, and another Wiener
process $w_-$ such that \begin{equation}\label{K3}
\xi(t)-\xi(s)=\int_s^t\gamma(\tau)d\tau+\sigma [w_-(t)-w_-(s)],\quad t_0\le
s<t\le t_1.
\end{equation}  Moreover, $\gamma$ also satisfies
\begin{equation}\label{K4}
E\left\{\int_{t_0}^{t_1}\gamma(t)\cdot\gamma(t)dt\right\}<\infty,
\end{equation} and $w_-(t)-w_-(s)$ is independent of $\{\xi(\tau);t\le \tau\le
t_1\}$. Let us agree that $dt$ always indicates a strictly positive variable.
For any function $f$ defined on
$[t_0,t_1]$, let

$$d_+f(t):=f(t+dt)-f(t)$$ be the {\it forward increment} at time $t$, and
$$d_-f(t)=f(t)-f(t-dt)$$ be the {\it backward increment} at time $t$. For a
finite-energy diffusion, F\"{o}llmer has also shown in \cite{F} that the forward
and backward drifts may be obtained as Nelson's conditional derivatives, namely
\begin{equation}\label{D1}\beta(t)=\lim_{dt\searrow
0}E\left\{\frac{d_+\xi(t)}{dt}|\xi(\tau), t_0 \le \tau \le
t\right\},\end{equation} and 
\begin{equation}\label{D2}\gamma(t)=\lim_{dt\searrow
0}E\left\{\frac{d_-\xi(t)}{dt}|\xi(\tau), t
\le  \tau \le t_1\right\},\end{equation} the limits being taken in
$L^2_n(\Omega,{\cal B},P)$. It was finally shown in \cite{F} that the one-time
probability density $\rho(\cdot,t)$ of $\xi(t)$ (which exists for every $t>t_0$)
is absolutely continuous on $\R^n$, and the following relation holds a.s.
$\forall t>0$ \begin{equation}\label{K4'} E\{\beta(t)-\gamma(t)|\xi(t)\} =
\sigma^2\nabla\log\rho(\xi(t),t). \end{equation} Let  $\xi$ be a finite-energy
diffusion satisfying (\ref{K1}) and (\ref{K3}). Let $f:\R^n\times [t_0,t_1]
\rightarrow \R$ be twice continuously differentiable with respect to the spatial
variable and once with respect to time. Then, we have the following change of
variables formulas: 
\begin{eqnarray}
\nonumber f(\xi(t),t)-f(\xi(s),s)=\int_s^t\left(\frac{\partial}{\partial
\tau}+\beta(\tau)\cdot\nabla+\frac{\sigma^2}{2}\Delta\right)f(\xi(\tau),\tau
)d\tau\\+\int_s^t\sigma\nabla f(\xi(\tau),\tau)\cdot
d_+w_+(\tau),\label{K5}\\\nonumber f(\xi(t),t)-f(\xi(s),s)=
\int_s^t\left(\frac{\partial}{\partial
\tau}+\gamma(\tau)\cdot\nabla-\frac{\sigma^2}{2}\Delta\right)f(\xi(\tau),\tau)d\tau
\\+\int_s^t\sigma\nabla f(\xi(\tau),\tau)\cdot d_-w_-(\tau). \label{K6}
\end{eqnarray} The stochastic integrals appearing in (\ref{K5}) and (\ref{K6})
are a (forward) Ito integral and a backward Ito integral, respectively, see
\cite {N3} for the details. Let us introduce the {\it current drift}
$v(t):=(\beta(t)+\gamma(t))/2$ and the {\it osmotic drift}
$u(t):=(\beta(t)-\gamma(t))/2$. Notice that, when $\sigma$ tends to zero, $v$
tends to
$\dot{\xi},$ and $u$ tends to zero. The semi-difference of (\ref{K1}) and
(\ref{K3}) gives the relation between the two driving ``noises" 
\begin{equation}0=\int_s^tu(\tau)d\tau+
\frac{\sigma}{2}\left[w_+(t)-w_+(s)-w_-(t)+w_-(s)\right].\label{K10} 
\end{equation}

The finite-energy diffusion $\xi(\cdot)$ is called {\it Markovian} if there
exist two measurable functions $b_+(\cdot,\cdot)$ and $b_-(\cdot,\cdot)$ such
that $\beta(t)=b_+(\xi(t),t)$ a.s. and
$\gamma(t)=b_-(\xi(t),t)$ a.s., for all $t$ in $[t_0,t_1]$. The duality relation
(\ref{K4'})  now reduces to Nelson's relation \cite{N58,N1} 
\begin{equation}\label{M1}  b_+(\xi(t),t)-b_-(\xi(t),t) =
\sigma^2\nabla\log\rho(\xi(t),t). 
\end{equation}  This immediately gives the {\it osmotic equation}
\begin{equation}\label{M2} u(x,t)=\frac{\sigma^2}{2}\nabla\log\rho(x,t),
\end{equation} where $u(x,t):=(b_+(x,t)-b_-(x,t))/2$. The probability density
$\rho(\cdot,\cdot)$ of $\xi(t)$ satisfies (at least weakly) the {\it
Fokker-Planck equation}
$$ \frac{\partial{\rho}}{\partial{t}} + \nabla \cdot (b_+\rho) =
\frac{\sigma^2}{2}\Delta\rho. $$  The latter can also be rewritten, in view of
(\ref{M1}), as the {\it equation of continuity} of hydrodynamics
\begin{equation}\label{M3}
\frac{\partial{\rho}}{\partial{t}} + \nabla \cdot (v\rho) = 0, \end{equation}
where
$v(x,t):=(b_+(x,t)+b_-(x,t))/2$.

\section{The quantum drift, the quantum noise and the bi-directional generator}

We recall now the basic facts from the time-symmetric kinematics employed in 
\cite{P1}-\cite{P4}. In order to develop stochastic mechanics as a
generalization of classical mechanics a salient difficulty is that the
finite-energy diffusion $\{x(t);t_0\le t\le t_1\}$ representing position of the
nonrelativistic particle has {\em two} natural velocities, namely the pair
($\beta(t),\gamma(t)$) or, equivalently, the pair ($v(t),u(t)$). It seems
therefore natural to replace the pair of real velocities by a unique
complex-valued  velocity. Since in the semiclassical limit we want to recover
the classical velocity, we only have the two choices $v\pm iu$. As observed in
\cite{P2}, $v-iu$ leads through a variational principle to the Schr\"{o}dinger
equation and $v+iu$ to the conjugate of the Schr\"{o}dinger equation,
respectively. For a general, finite-energy diffusion $\{\xi(t);t_0\le t\le
t_1\}$, how can we view the process
$v-iu$ as a drift? Let us multiply (\ref{K1}) by $\frac{1-i}{2}$ and (\ref{K3})
by
$\frac{1+i}{2}$, respectively, and then add. We get  \begin{eqnarray}\nonumber
&&\xi(t)-\xi(s)=\int_s^t\left[\frac{1-i}{2}\beta(\tau)+\frac{1+i}{2}\gamma(\tau)\right]d\tau\\&&
+\frac{\sigma}{2}\left[(1-i)(w_+(t)-w_+(s))+(1+i)(w_-(t)-w_-(s))\right].
\label{K11}\end{eqnarray}  We call
$$v_q(t):=\frac{1-i}{2}\beta(t)+\frac{1+i}{2}\gamma(t)=v(t)-iu(t)$$  the {\it
quantum drift}, and
\begin{equation}\label{K12}w_q(t):=\frac{1-i}{2}w_+(t)+\frac{1+i}{2}w_-(t)\end{equation}
the {\it quantum noise}. Hence, we can rewrite (\ref{K11}) as
\begin{equation}\label{K13}
\xi(t)-\xi(s)=\int_s^tv_q(\tau)d\tau+\sigma[w_q(t)-w_q(s)]. \end{equation} 
Representation (\ref{K13}) enjoys the
time reversal invariance property \cite{P2}. It
has been  employed in \cite{P1}-\cite{P3} in order to develop elements of
Lagrangian and Hamiltonian dynamics  in the frame of Nelson's stochastic
mechanics. In particular, to derive the second form of Hamilton's principle, the
key tool has been a change of variables formula related to representation
(\ref{K13}). In order to recall such a formula, we need first to define
stochastic integrals with respect to the quantum noise
$w_q$. Let us denote by
$$d_bf(t):=\frac{1-i}{2}d_+f(t)+\frac{1+i}{2}d_-f(t)$$  the  {\it bilateral
increment} of $f$ at time $t$. From (\ref{K12}) and (\ref{K10}), we get 
\begin{eqnarray}\label{FQN}d_+w_q(t)&=&\frac{1+i}{\sigma}u(t)dt+ d_+w_+ +o(dt),\\
d_-w_q(t)&=&\frac{-1+i}{\sigma}u(t)dt+ d_+w_- +o(dt).\label{BQN}
\end{eqnarray} These in turn give immediately the important relation:
\begin{proposition}
\begin{equation}\label{QN}d_bw_q(t):=\frac{1-i}{2}d_+w_+(t)+\frac{1+i}{2}d_-w_-(t)
+o(dt). 
\end{equation}
\end{proposition}  Let $f(x,t)$ be a measurable, $\C^n$-valued function such
that $$P\left\{\omega:\int_0^Tf(\xi(t),t)\cdot\overline{f(\xi(t),t)}dt
<\infty\right\}=1. $$ In view of (\ref{QN}), we define  
$$\int_s^tf(\xi(\tau),\tau)\cdot
d_bw_q(\tau):=\frac{1-i}{2}\int_s^tf(\xi(\tau),\tau)\cdot
d_+w_+(\tau)+\frac{1+i}{2}\int_s^tf(\xi(\tau),\tau)\cdot d_-w_-(\tau).$$ Thus,
integration with respect to the bilateral increments of $w_q$ is defined through
a linear combination with complex coefficients of a forward and a backward Ito
integral. Let $f(x,t)$ be a complex-valued function with real and imaginary
parts of class $C^{2,1}$. Then, multiplying (\ref{K5}) by $\frac{1-i}{2}$ and
(\ref{K6}) by $\frac{1+i}{2}$, respectively, and then adding, we get the change
of variables formula
\begin{eqnarray} f(\xi(t),t)-f(\xi(s),s)&=&\int_s^t\left(\frac{\partial}{\partial
\tau}+v_q(\tau)\cdot\nabla-\frac{i\sigma^2}{2}\Delta\right)f(\xi(\tau),\tau)
d\tau\nonumber\\&+&\int_s^t\sigma\nabla f(\xi(\tau),\tau)\cdot
d_bw_q(\tau).\label{K16} \end{eqnarray} Rewriting (\ref{K5})-(\ref{K6}) in
differential form, and exploiting (\ref{QN}), we get the differential form of
(\ref{K16})
\begin{equation}\label{CV}d_bf(\xi(t),t)=\left(\frac{\partial}{\partial
t}+v_q(t)\cdot\nabla-\frac{i\sigma^2}{2}\Delta\right)f(\xi(t),t)dt+\sigma
\nabla f(\xi(t),t)\cdot d_bw_q(t)+o(dt).
\end{equation} Finally, specializing (\ref{CV}) to $f(x,t)=x$, we get the
differential form of (\ref{K13})
$$d_b\xi(t)=v_q(t)dt+\sigma d_bw_q(t)+o(dt).
$$
A few remarks are now in order. As it is apparent from (\ref{FQN})-(\ref{BQN}),
there are profound differences between the representations (\ref{K1})-(\ref{K3})
and representation (\ref{K13}) for the increments of $\xi$. 
\begin{itemize}
\item The distribution of the quantum noise $w_q$ depends on the stochastic
process
$\xi$; 
\item Let ${\cal F}_t^-$ and ${\cal F}_t^+$ denote the $\sigma$-fields induced by
the past $\{\xi(\tau);t_0\le \tau\le t\}$ and the future $\{\xi(\tau);t\le
\tau\le t_1\}$ of $\xi$, respectively. The quantum noise
$w_q$ is not a forward $\{{\cal F}_t^-\}$-martingale neither a reverse-time
$\{{\cal F}_t^+\}$-martingale; 
\item The quantum noise $w_q$ is not Markovian even when $\xi$ is Markovian.
\end{itemize} The increments of the quantum noise $w_q$ are, nevertheless,
adapted both to the increasing filtration
${\cal F}^-:=\{{\cal F}_t^-\}$, and to the decreasing filtration ${\cal
F}^+:=\{{\cal F}_t^+\})$. Moreover,
$w_q$ is {\it mean-forward differentiable} \cite{N1} with respect to the
filtration ${\cal F}^-$ and the corresponding mean-forward derivative is
$$(D_+^{{\cal F}^-}w_q)(t)=\lim _{dt\searrow 0}E\left\{\frac{d_+w_q(t)}{dt}|{\cal
F}_t^-\right\}=\frac{1+i}{\sigma}u(t).
$$ Similarly,
$w_q$ is {\it mean-backward differentiable} with respect to the filtration
${\cal F}^+$ and the corresponding mean-backward derivative is
$$(D_-^{{\cal F}^+}w_q)(t)=\lim _{dt\searrow 0}E\left\{\frac{d_-w_q(t)}{dt}|{\cal
F}_t^+\right\}=\frac{-1+i}{\sigma}u(t).
$$ We then have the following remarkable result.
\begin{proposition}The quantum drift of $w_q$ with respect to  $({\cal F}^-,{\cal
F}^+)$ is zero, i.e.
$$ v_q^ {({\cal F}^-,{\cal
F}^+)}(w_q)(t):=\frac{1-i}{2}(D_+^{{\cal F}^-}w_q)(t)+\frac{1+i}{2}(D_-^{{\cal
F}^+}w_q)(t)=0,\quad\forall t\in [t_0,t_1].
$$
\end{proposition} Observing that, for all $t\in [t_0,t_1]$, we have $(D_+^{{\cal
F}^-}w_+)(t)=0$ and 
$(D_-^{{\cal F}^+}w_-)(t)=0$, we see that that  there is in fact a deep analogy
between the three driving processes in the representations (\ref{K1}),
(\ref{K3}) and (\ref{K13}). It follows from this result and (\ref{QN}), that the
quantum noise for $w_q$ corresponding to the pair of filtrations
$({\cal F}^-,{\cal F}^+)$ is $w_q$ itself. From now on, we consider the case
where
$\{\xi(t);t_0\le t\le t_1\}$ is {\em Markovian}.  The analogy between the three
driving noise can then also be seen in the following result \cite{P5}.
\begin{proposition}
\begin{eqnarray}E\{d_bw_q(t)|\xi(t)\}&=&0\\\label{M4}
 E\{d_bw_q(t)d_bw_q(t)^T|\xi(t)\}&=&-iI_ndt.
 \end{eqnarray}
\end{proposition}
Now let $L_+$ and $L_-$, defined by $$L_+:=b_+\cdot\nabla
+\frac{\sigma^2}{2}\Delta,\quad L_-:=b_-\cdot\nabla -\frac{\sigma^2}{2}\Delta, $$
be the forward and the backward generator of $\xi$, respectively.  Then
\cite{N3}, for a scalar $f$ of class $C^2$ with compact support in $\R^n$, we
have
\begin{eqnarray}\label{GEN1}\lim_{dt\searrow
0}E\left\{\frac{d_+f(\xi(t))}{dt}|\xi(t)=x\right\}&=&[L_+ f](x),\\
\lim_{dt\searrow 0}E\left\{\frac{d_-f(\xi(t))}{dt}|\xi(t)=x\right\}&=&[L_-
f](x)
\label{GEN2}
\end{eqnarray}
Let
$C_b^2(\R^n;\C)$ denote the complex, twice continuously differentiable,
functions with compact support in $\R^n$. For $f\in C_b^2(\R^n;\C)$, in view of
(\ref{K16}), we define the {\it bi-directional generator} $L_b$ of
$\{\xi(t)\}$ by 
\begin{equation}\label{GBD}L_bf=v_q\cdot\nabla f-\frac{i\sigma^2}{2}\Delta
f=\left[\frac{1-i}{2}L_++\frac{1+i}{2}L_-\right]f,
\end{equation} where the quantum drift field is
$$v_q(x,t):=\frac{1-i}{2}b_+(x,t)+\frac{1+i}{2}b_-(x,t).$$ 
Motivation for this definition is provided also by the following result:
\begin{proposition}
$$\lim_{dt\searrow
0}E\left\{\frac{d_bf(\xi(t))}{dt} | \xi(t)=x\right\}=[L_b f](x).
$$
\end{proposition}
Notice that the
operator $L_b$ is completely different from the generator of the bi-directional
Markov semigroup $\tilde{L}$ in \cite[Section 2]{AMU}.

\section{Discussion}
We come now to a crucial
point. Consider the forward driving noise $w_+$ in (\ref{K1}). Strictly
speaking, $w_+$ is originally only defined as an-dimensional {\it Wiener
difference process} $w_+(s,t)$, see \cite[Chapter 11]{N1} and
\cite[Section 1]{N3}. It is namely a process such that $w_+(t,s)=-w_+(s,t)$,
$w_+(s,u)+w_+(u,t)=w_+(s,t)$, and $w_+(s,t)$ is Gaussian distributed with mean
zero and variance $I_n|s-t|$. Moreover, (the components of) $w_+(s,t)$ and
$w_+(u,v)$ are independent whenever $[s,t]$ and $[u,v]$ don't overlap. Of course,
$w_+(t):=w_+(t_0,t)$ is a standard Wiener process such that 
$w_+(s,t)=w_+(t)-w_+(s)$ and $w_+(t_0)=0$. The fact that $w_+(t_0)=0$ is
important. It makes so that the past $\sigma$-fields generated by $w_+$ and by
the increments of $w_+$ coincide. Similarly, we can define $w_-$ of (\ref{K3})
so that
$w_-(t_1)=0$. Hence,   the future $\sigma$-fields generated by $w_-$ and by the
increments of $w_-$ are made to coincide. Now let $f:\R^n\times [t_0,t_1]
\rightarrow \C$ be of class $C^{2,1}$. Then, we have:  \begin{eqnarray}\nonumber
f(w_+(t),t)-f(w_+(s),s)&=&\int_s^t\left(\frac{\partial}{\partial
\tau}+\frac{1}{2}\Delta\right)f(w_+(\tau),\tau )d\tau+\int_s^t\nabla
f(w_+(\tau),\tau)\cdot d_+w_+(\tau)\\
f(w_-(t),t)-f(w_-(s),s)&=&\int_s^t\left(\frac{\partial}{\partial
\tau}-\frac{1}{2}\Delta\right)f(w_-(\tau),\tau )d\tau+\int_s^t\nabla
f(w_-(\tau),\tau)\cdot d_-w_-(\tau).\nonumber
\end{eqnarray} Thus, the forward generator of $w_+$ is $\frac{1}{2}\Delta$, and
the backward generator of $w_-$ is $-\frac{1}{2}\Delta$. It would be nice if we
could argue along the same lines that, for $f\in C_b^2(\C^n;\C)$, the 
bi-directional generator of the quantum noise is the operator
$$\frac{1-i}{2}(\frac{1}{2}\Delta)+\frac{1+i}{2}(-\frac{1}{2}\Delta)=-\frac{i}{2}\Delta.$$ 
But this is not possible because of measurability problems. Let us see why.
Instead of definition (\ref{K12}), we could start by defining $w_q$ only as a
difference process by $$w_q(s,t):=\frac{1-i}{2}w_+(s,t)+\frac{1+i}{2}w_-(s,t). $$
For a difference process $\theta(s,t)$, we define $d_+\theta(t):=\theta(t,t+dt)$ 
and $d_-\theta(t):=\theta(t-dt,t)$. We can then derive as before formulas
(\ref{FQN})-(\ref{QN}). Then, we would need to define the quantum noise $w_q$ at
some time $\bar{t}$ so that the process
$w_q(t):=w_q(\bar{t})+w_q(\bar{t},t)$ is simultaneously adapted to the two
filtrations induced by its past and future increments. But this is clearly
impossible. Hence, an object such as $$\int_s^t\nabla f(w_q(\tau),\tau)\cdot
d_bw_q(\tau)=\frac{1-i}{2}\int_s^t\nabla f(w_q(\tau),\tau)\cdot
d_+w_q(\tau)+\frac{1+i}{2}\int_s^t\nabla f(w_q(\tau),\tau)\cdot d_-w_q(\tau)
$$ cannot be given a meaning, since at least one of the two Ito integrals in the
right-hand side cannot be defined.

\section{Stochastic mechanics}

Nelson's stochastic mechanics \cite{N0}-\cite{BCZ} may be based, since the
important paper by Guerra and Morato
\cite{GM}, on stochastic variational principles of hydrodynamic type. Other
versions  of the variational principle have been proposed in \cite{N2,N3,P1}.
The solution of the stochastic variational principle is anyway a finite-energy
Markov diffusion process $\{x(t);t_0\le t\le t_1\}$ with diffusion coefficient
$\frac{\hbar}{m}$ to which it is naturally associated a quantum evolution
$\{\psi(x,t);t_0\le t\le t_1\}$, namely a solution of the Schr\"{o}dinger
equation  \begin{equation}\label{S}
\frac{\partial{\psi}}{\partial{t}} = \frac{i\hbar}{2m}\Delta\psi -
\frac{i}{\hbar}V(x)\psi, \end{equation} such that
\begin{equation}\label{F1}
\int_{t_0}^{t_1}\int_{\R^n}\left[\nabla\psi(x,t)\cdot\overline{\nabla\psi(x,t)}\right]dxdt
<\infty. \end{equation} The probability density $\rho(\cdot,t)$ of $x(t)$
satisfies
$\rho(x,t):=|\psi(x,t)|^2$, and the quantum drift field is given by
\begin{equation}\label{QDR}v_q(x,t)=\frac{\hbar}{mi}\nabla\log\psi(x,t).
\end{equation} Conversely, given a solution of the  Schr\"{o}dinger equation
$\{\psi(x,t);t_0\le t\le t_1\}$ satisfying the finite action condition
(\ref{F1}), a probability measure $P$ may be constructed on path space under
which the coordinate process is a finite-energy Markov diffusion with quantum
drift as in (\ref{QDR}), cf.\cite{Car}, \cite [Chapter IV]{BCZ}.

\section{Relation between the bi-directional
generator and the Hamiltonian operator}
In order to establish the relation in the section title, we need first the
following elementary result. 
\begin{lemma}\label{lem} Let $a$ and $b$ be two complex numbers, and let
$V:\R^n\rightarrow
\R$ be a measurable function. Let
$u:\R^n\times [t_0,t_1]\rightarrow\C$ be a {\rm never vanishing} solution of the
p.d.e. 
\begin{equation}\label{pd1}
\frac{\partial u}{\partial{t}}=a\Delta u+bVu, 
\end{equation}  on $[t_0,t_1]$. Then
$\theta:=u\phi$ is another solution of (\ref{pd1}) on $[t_0,t_1]$ if and only if
$\phi$ satisfies on the same time interval 
\begin{equation}\label{pd2}
\frac{\partial\phi}{\partial{t}}=2a\nabla \log u\cdot\nabla\phi+a\Delta\phi.
\end{equation} 
\end{lemma} 
\proof We have the following chain of equalities
\begin{eqnarray} &&\frac{\partial (u\phi)}{\partial{t}}=\frac{\partial
u}{\partial{t}}\phi+u\frac{\partial\phi}{\partial{t}}=a\Delta
u\phi+bVu\phi+u\frac{\partial\phi}{\partial{t}}\nonumber\\&&=a\left(\Delta
u\phi+2\nabla
u\cdot\nabla\phi+u\Delta\phi\right)+bVu\phi+u\frac{\partial\phi}{\partial{t}}-2a\nabla
u\cdot\nabla\phi-au\Delta\phi\nonumber\\&&=a\Delta(u\phi)+bV(u\phi)+u\left(\frac{\partial\phi}{\partial{t}}
-2a\frac{\nabla u}{u}\cdot\nabla\phi-a\Delta\phi\right).\nonumber
\end{eqnarray} 
\qed
\begin{remark}
{\em We shall apply Lemma \ref{lem} to both the diffusion and the
quantum case. Particularly for the latter application, it would be desirable to
have a more general result where $u$ may vanish. In order to avoid obscuring ideas
with technicalities, we shall be content here with discussing the non singular
case. It appears quite feasible, however, that applying ideas and results of
Carlen and others, see \cite[Chapter IV]{BCZ} and references therein, some of these
applications may be suitably extended to the singular case.}
\end{remark}  
\begin{lemma}Let $u$ and $V$ be as in {\em Lemma \ref{lem}}. Let
$C_b^{2,1}(\R^n\times [t_0,t_1];\C)$ denote the complex-valued functions of
class $C^{2,1}$ with compact support in $\R^n\times [t_0,t_1]$. On this domain,
we consider the operators  $$A:=\frac{\partial}{\partial{t}}-a\Delta-bM_V,$$ 
where
$M_V$ denotes the operator of multiplication by the function $V$, and 
$$B:=\frac{\partial}{\partial{t}}-2a\nabla\log u\cdot\nabla-a\Delta.$$ 
Then, for $f\in C_b^{2,1}(\R^n\times
[t_0,t_1];\C)$, we have
\begin{equation}\label{ABO}Bf=M_{u^{-1}}AM_uf. 
\end{equation}
\end{lemma}
Let ${\cal L}_c^{2,1}$ denote the Hilbert space of complex-valued functions $f$
satisfying
$$\int_{t_0}^{t_1}||f||^2_{{\cal L}_c^2(\R^n)}dt<\infty
$$
\begin{theorem}\label{th2} Let $a=\frac{i\hbar}{2m}$ and $b=-\frac{i}{\hbar}$ in
{\em Lemma \ref{lem}} Let
$\{\psi(x,t);t_0\le t\le t_1\}$ be a never vanishing solution of the
Schr\"{o}dinger equation (\ref{S}) satisfying (\ref{F1}). Let $L_b$ denote the
bi-directional generator of the associated Nelson process as defined in 
(\ref{GBD}), and let
${\bf H}=-\frac{\hbar^2}{2m}\Delta + V(x)$ denote the quantum Hamiltonian
operator. We consider the operator 
$(\frac{\partial}{\partial{t}}+\frac{i}{\hbar}{\bf H})$ defined in 
${\cal L}_c^{2,1}$. Let
${\cal L}_c^{2,1}(|\psi|^2)$ denote the Hilbert space of functions $g$ such that
$(g\psi)\in {\cal L}_c^{2,1}$. Then,  $(\frac{\partial}{\partial{t}}+L_b)$
defined in ${\cal L}_c^{2,1}(|\psi|^2)$ and 
$(\frac{\partial}{\partial{t}}+\frac{i}{\hbar}{\bf H})$ are unitarily
equivalent.  Indeed, it follows from (\ref{ABO}) that 
\begin{equation}\label{ABOR}\frac{\partial}{\partial{t}}+L_b=M_{\psi}^{-1}
\left(\frac{\partial}{\partial{t}}+\frac{i}{\hbar}{\bf H}\right)M_{\psi}.
\end{equation} \end{theorem}

\begin{remark}{\em Relation (\ref{ABOR}) supports the choice of the kinematics
of Section 3 to study quantum-mechanical problems. It may be viewed as a
generalization of a well-known result relating the usual generator to
the Hamiltonian operator through the {\em ground state transformation}, see e.g.
\cite{AHG,AHGS,25}. Indeed, for $\psi(x,t)=\psi_0(x)$ the ground state of the
Hamiltonian (${\bf H}\psi_0=0$), and
$f\in {\cal L}^2(\R^n;|\psi_0|^2dx)$, (\ref{ABOR}) reads
$$\frac{\hbar}{im}\nabla\log\psi_0\cdot\nabla f-\frac{i\hbar}{2m}\Delta
f=\frac{i}{\hbar}M_{\psi_0}^{-1}{\bf H}M_{\psi_0} f.
$$ This immediately gives
\begin{equation}\label{ABORI}-\frac{\hbar^2}{m}\left(\nabla\log\psi_0\cdot\nabla
f+\frac{1}{2}\Delta f\right)=M_{\psi_0}^{-1}{\bf H}M_{\psi_0} f.
\end{equation}
\noindent }
\end{remark}

\section{The Feynman-Kac formula}
Let $h:\R^n\times [t_0,t_1]\rightarrow \R$ be a classical, never
vanishing solution of the terminal value problem
\begin{equation}\label{FK0}\frac{\partial h}{\partial t}+\frac{1}{2}\Delta
h=V(x)h,\quad h(x,t_1)=h_1(x),\end{equation}  where $V$ is a nonnegative,
measurable function on $\R^n$. A simple calculation shows that $\log h$ satisfies
\begin{equation}\label{FK1} \frac{\partial\log h}{\partial t}+\nabla\log
h\cdot\nabla\log h+\frac{1}{2}\Delta\log h=\frac{1}{2}\nabla\log
h\cdot\nabla\log h+V(x). \end{equation} Assume that there exists a weak solution
$P$ on $[t_0,t_1]$ of the stochastic differential equation $$dx=\nabla\log h
\;dt+ dw.$$ 
Namely, the coordinate process 
$\{x(t);t_0\le t\le t_1\}$ under $P$ admits the above forward differential.
Applying Lemma \ref{lem} to the diffusion case, we get a different
generalization of (\ref{ABORI}). Let ${\cal L}^{2,1}$ denote the Hilbert space 
of real-valued functions $f$
satisfying
$$\int_{t_0}^{t_1}||f||^2_{{\cal L}^2(\R^n)}dt<\infty
$$
\begin{proposition} Let $a=-\frac{1}{2}$ and $b=1$ in {\em Lemma \ref{lem}}. Let
$h(x,t);t_0\le t\le t_1\}$ be a never vanishing solution of  equation
(\ref{FK0}). Let 
$$L_+=\nabla\log h\cdot\nabla+\frac{1}{2}\Delta$$  denote the generator of the
measure $P$ and let $H=-\frac{1}{2}\Delta + V(x)$ denote the Hamiltonian
operator. We consider the operator 
$(\frac{\partial}{\partial{t}}-H)$ defined in 
${\cal L}^{2,1}$.  Let
${\cal L}^{2,1}(h^2)$ denote the Hilbert space of functions $g$ such that
$(gh)\in {\cal L}^{2,1}$.  Then,  $(\frac{\partial}{\partial{t}}+L_+)$
defined in ${\cal L}^{2,1}(h^2)$ and 
$(\frac{\partial}{\partial{t}}-H)$ are unitarily
equivalent. Indeed, it follows from (\ref{ABO}), that 
\begin{equation}\label{ABORR}
\frac{\partial}{\partial{t}}+L_+=M_{h}^{-1}
\left(\frac{\partial}{\partial{t}}-H\right)M_{h}.
\end{equation}
\end{proposition}
\noindent
We recall below three derivations of the Feynman-Kac formula, see e.g.
\cite{25}. These will serve for the purpose of comparison in the following
section. Hence, no effort will be made for maximal generality.\\
\noindent
{\em Derivation 1.}\\
\noindent
Suppose now that, under $P$, $\{x(t)\}$ is a finite energy diffusion. Under
$P$, we have
$$h(x(t),t)=h_1(x(t_1))\exp[-\int_t^{t_1}d\log h(x(\tau),\tau)d\tau].$$ By Ito's
rule, and (\ref{FK1}), we get 
\begin{equation}\label{FK2}
h(x(t),t)=h_1(x(t_1))\exp\left\{-\int_t^{t_1}\left[\frac{1}{2}\nabla\log h\cdot
\nabla\log h+V\right]d\tau-\int_t^{t_1}\nabla\log h\cdot dw(\tau)\right\}.
\end{equation}  Let us introduce the random variable
\begin{eqnarray}\nonumber
Z_t^{t_1}&=&\exp\left\{-\int_t^{t_1}\frac{1}{2}\nabla\log h\cdot \nabla\log h
d\tau-\int_t^{t_1}\nabla\log h\cdot
dw(\tau)\right\}\\&=&\exp\left\{\int_t^{t_1}\frac{1}{2}\nabla\log h\cdot
\nabla\log h d\tau-\int_t^{t_1}\nabla\log h\cdot dx(\tau)\right\},\label{FK3}
\end{eqnarray} and rewrite (\ref{FK2}) as  \begin{equation}\label{FK2'}
h(x(t),t)=h_1(x(t_1))\exp\left\{-\int_t^{t_1}V(x(\tau))d\tau\right\}Z_t^{t_1}.
\end{equation} Now let $P_{tx}$ denote the conditional measure
$P[\cdot|x(t)=x]$. Integrating on both sides of (\ref{FK2'}) with respect to
$P_{tx}$, we get  
\begin{equation}\label{FK4}
h(x,t)=\int_{\Omega}h_1(x(t_1))\exp\left\{-\int_t^{t_1}V(x(\tau)
d\tau\right\}Z_t^{t_1}dP_{tx},  \end{equation} where $\Omega=C([t,t_1];\R^n)$.
By the finite energy assumption, $Z_t^{t_1}$ induces a measure transformation
\cite{KS}. Indeed,
$dW_{tx}=Z_t^{t_1}dP_{tx}$, where $W_{tx}$ denotes Wiener measure starting at
$x$ at time $t$. Hence, (\ref{FK4}) gives
\begin{equation}\label{FK5}
h(x,t)=\int_{\Omega}h_1(\omega(t_1))\exp\left\{-\int_t^{t_1}V(\omega(\tau)
d\tau\right\}dW_{tx}(\omega), \end{equation} which is the Feynman-Kac formula.
The above derivation of (\ref{FK5}), based on the Girsanov transformation, is by
no means the simplest.  The simplest derivation of (\ref{FK5}) is, in our
opinion, the following. 
\\ {\em Derivation 2.}\\
\noindent Let $\{w(\tau):t\le \tau\le t_1\}$ be a standard, n-dimensional Wiener
process such that $w(t)=x$. Let us introduce the process
$y(\tau):=h(w(\tau),\tau)$. By Ito's rule, and equation (\ref{FK0}), we have
\begin{equation}\label{FK6} dy=V(w(\tau))y(\tau)d\tau+\nabla
h(w(\tau),\tau)\cdot dw.
\end{equation} The crucial observation here is that $y$ satisfies a {\it linear}
stochastic differential equation (with random, but adapted to the past of $w$,
coefficient $V(w(\tau))$. It is natural to try to solve the equation with the
aid of an integrating factor. We multiply both sides of (\ref{FK6}) by
$\exp \left(-\int_t^\tau V(w(\sigma))d\sigma\right)$ and get \begin{equation}
d\left[\exp \left(-\int_t^\tau V(w(\sigma))d\sigma\right)y\right]=\exp
\left(-\int_t^\tau V(w(\sigma))d\sigma\right)\nabla h(w(\tau),\tau)\cdot dw
\end{equation} Integrating between
$t$ and $t_1$, we get
\begin{equation} \exp \left(-\int_t^{t_1} V(w(\sigma))d\sigma\right)y(T)-y(t)=
\int_t^{t_1}\exp
\left(-\int_t^\tau V(w(\sigma))d\sigma\right)\nabla h(w(\tau),\tau)\cdot dw.
\end{equation} Let us assume that
$$E\left\{\int_{t_0}^{t_1}\nabla h(w(\tau),\tau)\cdot\nabla
h(w(\tau),\tau)d\tau\right\}<\infty.
$$ Then, observing that $\exp\left(-\int_t^\tau V(w(\sigma))d\sigma\right)$ is
bounded, we conclude that the stochastic integral on the right-hand side is a
martingale. Taking the conditional expectation $E\{\cdot|w(t)=x\}$ on both
sides, we get (\ref{FK5}).
\\ {\em Derivation 3.}\\ \noindent We shall now look at the derivation of the
Feynman-Kac formula based on the Trotter product formula.
We consider first the case
$V\equiv 0$. Let $q(t,x,t_1,y)$ be the transition density of the measure $P$.
Taking $a=-\frac{1}{2}$ in Lemma \ref{lem}, we get that 
$$\frac{h(x,t)}{h_1(y)}q(t,x,t_1,y) $$ is the fundamental solution of
$$\frac{\partial u}{\partial{t}}+\frac{1}{2}\Delta u=0. $$  
\begin{proposition}\label{pro7}
The kernel
$$\frac{h(x,t)}{h_1(y)}q(t,x,t_1,y)
$$ does not depend on $\{h(x,t);t_0\le t\le t_1\}$. Indeed,
\begin{equation}\label{pro70}
\frac{h(x,t)}{h_1(y)}q(t,x,t_1,y)=p(t,x,t_1,y)=[2\pi(t_1-t)]^{-\frac{n}{2}}\exp
\left[-\frac{|x-y|^2}{2(t_1-t)}\right]. \end{equation} 
\end{proposition} Notice that relation (\ref{pro70}) between transition
densities mirrors the corresponding relation between probability measures that,
in view of (\ref{FK2'}), here reads
$$\frac{h(x(t),t)}{h_1(x(t_1))}dP_{tx}=dW_{tx}. $$ From (\ref{pro70}), we
immediately get
$$ h(x,t)=E\{h_1(w(t_1)|w(t)=x\}=\int_{\Omega}h_1(\omega(t_1))dW_{tx}(\omega).
$$ Consider now the case where $V$ is any continuous function. An interesting
consequence of Lemma \ref{lem} is the following. Let 
$\{h_2(x,\tau);t_0\le t\le t_1\}$ be another solution of (\ref{FK0}). Let
$$\varphi(x,t):=\frac{h_2(x,t)}{h(x,t)}.
$$
\begin{corollary}\label{cor1} Under $P$, the stochastic process
$\varphi(x(t),t)$ satisfies
\begin{equation}\label{LM}\varphi(x(t),t)-\varphi(x(s),s)=\int_s^t\nabla\varphi(x(\tau),\tau)\cdot
dw(\tau),\quad s<t. 
\end{equation}
\end{corollary} 
\proof By Lemma \ref{lem}, 
$$\left[\frac{\partial}{\partial t}+\nabla\log
h(x,t)\cdot\nabla+\frac{1}{2}\Delta\right] \varphi=0.
$$ By Ito's rule, we now get (\ref{LM}).
\qed
\noindent Now let
$q(t,x,t_1,y)$ be the transition density of the measure $P$. Taking
$a=-\frac{1}{2}$ and $b=1$ in the Lemma \ref{lem},  we get that $w(t,x,t_1,y)$
defined by
$$w(t,x,t_1,y):=\frac{h(t,x)}{h_1(y)}q(t,x,t_1,y)$$  is another solution of
equation (\ref{FK0}). Let us find some heuristic connection between
$w(t,x,t_1,y)$ and the kernel $p(t,x,t_1,y)$ in (\ref{3}). Let
$$r(t,x,t_1,y,x_1):=w(t,x,t_1,y)\exp [V(x_1)(t_1-t)].$$ 
$r$ satisfies
$$\frac{\partial r}{\partial t}+\frac{1}{2}\Delta r=[V(x)-V(x_1)]r.
$$ Then, for $|x_1-x|$ small, the function
$r(t,x,t_1,y,x_1)$ is close to $p(t,x,t_1,y)$. Now let $\omega(\cdot)$ be a
continuous curve on $[t,t_1]$, and let
$x_j=\omega(t+(t_1-t)j/l), j=0,1,\ldots,l.$ Iterating, we then get
\begin{eqnarray}\nonumber
h(t,x)&&=\lim_{l\rightarrow\infty}[2\pi(t_1-t)/l]^{-\frac{nl}{2}}\\&&\int\cdots\int\exp\left[-\sum_{j=1}^{l}
\frac{t_1-t}{l}\left(-\frac{|x_j-x_{j-1}|^2}
{2(t_1-t)/l}+V(x_j)\right)\right]h(t_1,x_l)dx_l\cdots dx_1
.\label{FK7}\end{eqnarray} Observing that 
$$\int_t^{t_1}-\frac{1}{2}\dot{\omega}(\tau)^2 d\tau
$$ may be viewed as the density of Wiener measure with respect to a (fictitious)
uniform measure on $\R^\infty$, we recognize that (\ref{FK7}) coincides with the
Feynman-Kac formula (\ref{FK5}).  For $V$ in the Kato class, this heuristic
argument can be turned into the rigorous one of Theorem \ref{th1} by means of
the Trotter formula \cite{N64}.

\section{Feynman integrals}

Let $\{\psi(x,t);t_0\le t\le t_1\}$ be the solution of the Schr\"{o}dinger
equation (\ref{S}) with initial condition
$\psi(x,t_0)=\psi_0(x)$. We suppose that $\psi$  never vanishes and satisfies 
\begin{equation}\label{F1o}
\int_{t_0}^{t_1}\int_{\R^n}\left[\nabla\psi(x,t)\cdot\overline{\nabla\psi(x,t)}\right]dxdt
<\infty. 
\end{equation}  Hence, the finite energy condition of \cite{Car} is satisfied, and
there exists a probability measure $P$ on path space under which the coordinate
process has forward drift field 
$$v(x,t)+u(x,t)=\frac{\hbar}{m}\nabla\left[\Im 
\log\psi(x,t)+\Re\log \psi(x,t)\right],$$ and quantum drift field $v_q(x,t)=
\frac{\hbar}{mi}\nabla\log\psi(x,t)$. Let $\{x(t);t_0\le t\le t_1\}$ denote the
coordinate process with the Nelson measure $P$. Observe that
$\log\psi(x,t)$ satisfies \begin{equation}\label{F2} \frac{\partial
\log\psi}{\partial{t}} + \frac{\hbar}{2im}\nabla \log\psi \cdot \nabla\log\psi +
\frac{i}{\hbar}V(x) - \frac{i\hbar}{2m}\Delta\log\psi = 0. \end{equation}

We now seek to derive a path-integral representation for $\psi(x,t)$ adapting to
the present setting the first derivation of the Feynman-Kac formula in the
previous section. Under the Nelson measure
$P$, we have \begin{equation}\label{F3}
\psi(x(t),t)=\psi_0(x(0))\exp\left[\log\psi(x(t),t)-\log\psi_0(x(0))\right].
\end{equation} By the change of variables formula (\ref{K16}), we get
\begin{eqnarray}\nonumber
&&\psi(x(t),t)=\psi_0(x(0))\times\\&&\exp\left\{\int_0^t\left[\frac{\partial}{\partial
\tau}+[v(x(\tau),\tau)-iu(x(\tau),\tau)]\cdot\nabla
-\frac{i\hbar}{2m}\Delta\right]\log\psi(x(\tau),\tau)d\tau+\right.\nonumber\\&&\left.\int_0^t\nabla\log\psi(x(\tau),\tau)\cdot
d_bw_q(\tau)\right\}.\label{F4} \end{eqnarray} By equation (\ref{F2}), and
recalling that
$$v(x(\tau),\tau)-iu(x(\tau),\tau)=\frac{\hbar}{im}\nabla\log\psi(x(\tau),\tau),$$ 
we get \begin{eqnarray}\nonumber
&&\psi(x(t),t)=\psi_0(x(0))\times\\&&\exp\left\{\int_0^t\left[\frac{im}{2\hbar}[v(x(\tau),\tau)-iu(x(\tau)]\cdot
[v(x(\tau),\tau)-iu(x(\tau)]-\frac{i}{\hbar}V(x(\tau)\right]\right.\nonumber\\&&\left.\int_0^t\nabla\log\psi(x(\tau),\tau)\cdot
d_bw_q(\tau)\right\}.\label{F5} \end{eqnarray} Let us introduce the random
variable \begin{eqnarray}\nonumber
\tilde{Z}_{t_0}^{t}:=&&\exp\left\{\int_0^t\frac{im}{2\hbar}[v(x(\tau),\tau)-iu(x(\tau),\tau)]\cdot
[v(x(\tau),\tau)-iu(x(\tau),\tau)]d\tau\right.\\&&\left.+
\int_0^t\frac{im}{\hbar}[v(x(\tau),\tau)-iu(x(\tau),\tau)]\cdot
d_bw_q(\tau)\right\}\nonumber\\&&=\exp\left\{\int_0^t\frac{-im}{2\hbar}[v(x(\tau),\tau)-iu(x(\tau),\tau)]\cdot
[v(x(\tau),\tau)-iu(x(\tau),\tau)]d\tau\right.\nonumber\\&&\left.
+\int_0^t\frac{im}{\hbar}[v(x(\tau),\tau)-iu(x(\tau),\tau)]\cdot 
d_bx(\tau)\right\},\label{F6} \end{eqnarray} and rewrite (\ref{F5}) as
\begin{equation}\label{F7}
\psi(x(t),t)=\psi_0(x(0))\exp\left\{\int_0^t\left[-\frac{i}{\hbar}V(x(\tau)\right]
d\tau\right\}\tilde{Z}_{t_0}^{t}.
\end{equation} Let $P_{tx}$ denote the conditional Nelson measure
$P[\cdot|x(t)=x]$ on $\Omega=C([0,t],\R^n)$. Taking expectations of both sides
of (\ref{F7}) with respect to $P_{tx}$, we get
\begin{equation}\label{F8}
\psi(x,t)=\int_\Omega\psi_0(\omega(0))\exp\left\{\int_0^t\left[-\frac{i}{\hbar}
V(\omega(\tau)\right]d\tau\right\}\tilde{Z}_{t_0}^{t}dP_{tx}(\omega).
\end{equation} This representation appears similar to representation (\ref{FK4})
for the solution
$h(x,t)$ of the antiparabolic equation of the previous section.  What made
(\ref{FK4}) useful was the relation  $Z_t^{t_1}dP_{tx}=dW_{tx}$ showing that the
product
$Z_t^{t_1}dP_{tx}$ is a universal measure on path space {\it independent of the
particular solution} $h(x,t)$.  It is apparent that $\tilde{Z}_{t_0}^{t}$ cannot
be a Radon-Nikodym derivative between two probability measures on path space
since it is {\it complex-valued}.  We are then led to the following two crucial
questions: 
\begin{enumerate} 
\item Is
$\tilde{Z}_{t_0}^{t}dP_{tx}$ a {\it bona fide} complex measure of bounded
variation (see Appendix A) on $C([0,T];\R^n)$? 
\item Is $\tilde{Z}_{t_0}^{t}dP_{tx}$ in some appropriate sense independent from
the particular solution $\{\psi(x,t)\}$, i.e. is it independent of $\psi_0(x)$
and of $V(x)$?  
\end{enumerate}  Obviously, we expect a negative answer to the second question
as the quantum noise, to which the``measure" $\tilde{Z}_{t_0}^{t}dP_{tx}$ should
correspond, does depend on the particular solution  $\{\psi(x,t)\}$. 
\begin{proposition}\label{pro} Let  $\{\psi(x,t);t_0\le t\le t_1\}$ be a never
vanishing solution of the Schr\"{o}dinger equation (\ref{S}) with initial
condition
$\psi(x,t_0)=\psi_0(x)$, and satisfying (\ref{F1o}). Assume that $\psi_0\in
L^1(\R^n)$. Let
$P_{tx}$ be the conditional Nelson measure associated to $\{\psi(x,s);0\le s\le
t\}$ and let
$\tilde{Z}_{t_0}^{t}$ be defined by (\ref{F6}). Then, $\tilde{Z}_{t_0}^{t}\in
L^1(P_{tx})$. It follows that
$d\mu:=\tilde{Z}_{t_0}^{t}dP_{tx}$ is a complex measure of bounded variation on
$C([t_0,t];\R^n)$.
\end{proposition} \proof   Taking absolute values on both sides of (\ref{F7}),
and recalling Born's relation
$|\psi(x,\tau)|^2=\rho(x,t)$ relating the wave function to the probability
density of the Nelson process at time $t$, we get
\begin{equation}\label{TV} |\tilde{Z}_{t_0}^{t}|=
\frac{\rho^{1/2}(x(t),t)}{\rho_0^{1/2}(x(0))}\,
\end{equation} where
$\rho_0(x)=|\psi_0(x)|^2$. Hence
$$\int|\tilde{Z}_{t_0}^{t}|dP_{tx}=\int\frac{\rho^{1/2}(x(t),t)}{\rho_0^{1/2}(x(0))}dP_{tx}=
\rho^{1/2}(x,t)\int_{\R^n}\rho_0^{1/2}(x)dx=\rho^{1/2}(x,t)\int_{\R^n}|\psi_0(x)|dx<\infty$$
\qed 
\noindent Thus, under the hypothesis and in the notation of the above
proposition, we can rewrite (\ref{F8}) in the form
\begin{equation}\label{F8'}
\psi(x,t)=\int_\Omega\psi_0(\omega(0))\exp\left\{\int_0^t\left[-\frac{i}{\hbar}
V(\omega(\tau)\right]d\tau\right\}d\mu(\omega). \end{equation} It follows,
however, from (\ref{TV}) that the {\em total variation} $|\mu|$ of
$\mu$ satisfies
$$d|\mu|(\omega)=|\tilde{Z}_{t_0}^{t}|dP_{tx}(\omega)=
\frac{\rho^{1/2}(x,t)}{\rho_0^{1/2}(\omega(0))}dP_{tx}(\omega).
$$ Thus, the measure $\mu$ does depend on the particular solution
$\{\psi(x,t)\}$. An attempt to derive a path-integral representation for
$\psi(x,t)$ along the lines of the second derivation of the Feynman-Kac formula
appears hopeless because $\psi(w_q(t),t)$ makes no sense since $w_q$ has complex
values and, more importantly, because of the considerations made in
Section 4. We turn, therefore, to the third derivation. Consider first the case
$V=0$. In view of the change of variable formula (\ref{CV}), we take
$p_q(t_0,y,t,x)$ to be the fundamental solution of the equation 
\begin{equation}\label{BDG}\left(\frac{\partial}{\partial
t}+v_q(x,t)\cdot\nabla-\frac{i\hbar}{2m}\Delta\right)u=0,
\end{equation} where, as usual,
$v_q(x,t)=\frac{\hbar}{im}\nabla\log\psi(x,t)$. Taking
$a=\frac{i\hbar}{2m}$ in Lemma \ref{lem}, we get that 
$$\frac{\psi(t,x)}{\psi_0(y)}p_q(t_0,y,t,x) $$ is the fundamental solution of
\begin{equation}\label{FS}\frac{\partial u}{\partial{t}}-\frac{i\hbar}{2m}\Delta
u=0. 
\end{equation} Hence, we get the counterpart of Proposition \ref{pro7}.
\begin{proposition}\label{pro2} The kernel
$$\frac{\psi(t,x)}{\psi_0(y)}p_q(t_0,y,t,x)
$$ does not depend on $\{\psi(x,t);t_0\le t\le t_1\}$. Indeed,
\begin{equation}\label{F9}
\frac{\psi(t,x)}{\psi_0(y)}p_q(t_0,y,t,x)=K(t_0,y,t,x)=\left[\frac{2\pi\hbar
i(t-t_0)}{m}\right]^{-\frac{n}{2}}\exp
\left[\frac{im|x-y|^2}{2\hbar(t-t_0)}\right]. \end{equation}
\end{proposition} Consider now the case where $V$ is any continuous function.
Let $\{\psi(x,t);t_0\le t\le t_1\}$ be a never vanishing solution of the
Schr\"{o}dinger equation (\ref{S}) with initial condition
$\psi(x,t_0)=\psi_0(x)$, and satisfying (\ref{F1o}), and let 
$\{\psi_2(x,t);t_0\le t\le t_1\}$ be another solution of (\ref{S}). Let
$$\tilde{\varphi}(x,t):=\frac{\psi_2(x,t)}{\psi(x,t)}.
$$
\begin{corollary}Under the Nelson measure $P$ associated to
$\{\psi(x,t);t_0\le t\le t_1\}$, the stochastic process $\tilde{\varphi}(x(t),t)$
satisfies
\begin{equation}\label{LM2}\tilde{\varphi}(x(t),t)-\tilde{\varphi}(x(s),s)=
\int_s^t\sqrt{\frac{\hbar}{m}}\nabla\tilde{\varphi}(x(\tau),\tau)\cdot
d_bw_q(\tau),\quad s<t.  \end{equation}
\end{corollary} 
\proof By Lemma \ref{lem}, 
\begin{equation}\label{CWF}\left[\frac{\partial}{\partial
t}+v_q(x,t)\cdot\nabla-\frac{i\hbar}{2m}\Delta\right] \tilde{\varphi}=0,
\end{equation} 
where $v_q(x,t)=\frac{\hbar}{im}\nabla\log\psi(x(t)$. By
(\ref{K16}), we now get (\ref{LM2}).
\qed
\noindent This result is the counterpart of Corollary \ref{cor1}. Notice that,
since the ratio of two solutions of the Schr\"{o}dinger equation satisfies
(\ref{CWF}), the function 
$$\theta(x,t):=\log\frac{\psi_2(x,t)}{\psi(x,t)}$$
satisfies the nonlinear equation
$$\frac{\partial{\theta}}{\partial{t}}
+ v_q(x,t)\cdot \nabla \theta(x,t)-\frac{i\hbar}{2m}\Delta
\theta(x,t)=\frac{i\hbar}{2m}\nabla\theta(x,t)\cdot\nabla\theta(x,t).
$$
This is precisely the Hamilton-Jacobi-type equation associated to the
variational problem that produces the new Nelson process after a position
measurement, causing the ``collapse of the wave function", see
\cite[Section VI]{P4}.\\ 
\noindent
Now, let $p_q(t_0,y,t,x)$  be the fundamental solution
of (\ref{BDG}). Taking $a=\frac{i\hbar}{2m}$ and $b=-\frac{i}{\hbar}$ in Lemma
\ref{lem}, we get that $\tilde{w}(t_0,y,t,x)$ defined by
$$\tilde{w}(t_0,y,t,x):=\frac{\psi(t,x)}{\psi_0(y)}p_q(t_0,y,t,x)$$  is
another solution of the Schr\"{o}dinger equation (\ref{S}). Let us find some
heuristic connection between $\tilde{w}(t_0,y,t,x)$ and the kernel
$K(t_0,y,t,x)$ in (\ref{1}). Let
$\tilde{r}(t_0,y,t,x,x_1):=\tilde{w}(t_0,y,t,x)\exp
[\frac{i}{\hbar}V(x_1)(t-t_0)]$. Then $\tilde{r}$ satisfies $$\frac{\partial
\tilde{r}}{\partial t}-\frac{i\hbar}{2m}\Delta
\tilde{r}=\frac{i}{\hbar}[V(x_1)-V(x)]\tilde{r}. $$ For $|x_1-x|$ small, the
function $\tilde{r}(t_0,y,t,x,x_1)$ is close to $K(t_0,y,t,x)$.   Now let
$\omega(\cdot)$ be a continuous curve on $[t_0,t]$, and let
$x_j=\omega(t_0+(t-t_0)j/l), j=0,1,\ldots,l.$ Iterating, we then get
\begin{eqnarray}\nonumber
\psi(t,x)&=&\lim_{l\rightarrow\infty}\left[\frac{2\pi\hbar
i(t-t_0)}{lm}\right]^{-\frac{nl}{2}}\\&&\int\cdots\int\exp\left[-\sum_{j=1}^{l}
\frac{i}{\hbar}\left(-\frac{m}{2\hbar}\frac{|x_j-x_{j-1}|^2}
{(t-t_0)/l}+V(x_j)\frac{(t-t_0)}{l}\right)\right]\psi_0(x_l)dx_l\cdots dx_1
.\nonumber \end{eqnarray}  This heuristics can be turned into the rigorous
argument of Theorem \ref{th1} by means of the Kato-Trotter formula \cite{N64}.

\section{Closing comments}

We have shown that, employing the time-symmetric kinematics of Section 3, it is
possible to establish a link between Nelson's stochastic mechanics and the
Feynman integral. Not surprisingly, we do have  the following negative result. It
is not possible to view the operator
$-\frac{i}{2}\Delta$ as the bi-directional generator of the quantum noise, as
argued in Section 4. Moreover, the complex measure $\mu$ in Proposition
\ref{pro} does depend on the particular solution of the Schr\"{o}dinger
equation. Nevertheless, the results in the second part of the previous section
show that the analogy with the diffusion case goes far beyond what was
believed, provided the time-symmetric kinematics of stochastic mechanics is
employed in the quantum case.

In \cite{G}, concerning the Feynman integral and stochastic mechanics, Guerra
writes: ``The full clarification of the deep connection between the two
approaches will be a major step toward a better understanding of the physical
foundations of quantum mechanics". We hope that this paper will stimulate new
research in this direction.

\noindent{\bf Acknowledgment} 

I wish to thank Paolo Dai Pra for several valuable discussions.

\appendix  \section{Complex measures} We collect in this appendix a few basic
facts about complex measures. We refer the reader to \cite[Chapter 6]{Rudin} for
the proofs and more information.

Let $\Omega$ be a set and $\cal{B}$ a $\sigma$-algebra of subsets of
$\Omega$. A complex function $\mu$ on $\cal{B}$, i.e. $\mu:
\cal{B}\rightarrow\C$, is called a {\it complex measure} on $\cal{B}$ if, for
every $B\in \cal{B}$, $$\mu(B)=\sum_{i=1}^\infty\mu(B_i)$$ holds whenever
$\{B_i\}_{i=1}^\infty$ is a countable partition of the set $B$. It is implicit
in this definition that every such series must converge.

Let $\mu$ be a complex measure. Then, among all {\it positive}, i.e. usual,
measures $\lambda$ satisfying $|\mu(B)|\le \lambda(B), \forall B\in
\cal{B}$, there exists a least one called {\it total variation} of $\mu$ and
denoted by $|\mu|$. The measure $|\mu|$ is minimal among all positive measures
$\lambda$ described above in the sense that $|\mu|(B)\le
\lambda(B)$ for all $B\in \cal{B}$. The measure $|\mu|$ has the remarkable
property that $|\mu(\Omega)| <\infty$. Thus the range of every complex measure
$\mu$ lies in a disc of finite radius. It is then usual to say that
$\mu$ is {\it of bounded variation}. \begin{theorem} Let $\lambda$ be a
positive, $\sigma$-finite measure on $\cal{B}$. Let $\mu$ be a complex measure
on $\cal{B}$. Suppose that $\mu$ is {\it absolutely continuous} with respect to
$\lambda$, namely $\mu(B)=0$ for every $B\in \cal{B}$ for which $\lambda(B)=0$.
Then there exists a unique function $h\in L^1(\lambda)$ such that
$$\mu(B)=\int_Bhd\lambda$$ for every $B\in
\cal{B}$. \end{theorem} A consequence of this theorem taking
$\lambda=|\mu|$, is the following result. \begin{theorem}Let $\mu$ be a complex
measure on $\cal{B}$. Then there exists a unimodular function $h$, i.e.
$|h(\omega)|=1$ for all $\omega\in \Omega$, such that the following {\it polar
decomposition} of $\mu$ holds $$d\mu=h\,d|\mu|.$$ \end{theorem} We also have the
following result. \begin{theorem} Suppose $\lambda$ is a positive measure on
$\cal{B}$, $h\in L^1(\lambda)$, and $\mu$ is the complex measure on
$\cal{B}$ defined by $$\mu(B)=\int_Bhd\lambda.$$ Then, for all $B\in \cal{B}$,
we have $$|\mu|(B)=\int_B|h|d\lambda.$$
\end{theorem}

\end{document}